# Thermal balance of tungsten monocrystalline nanoparticles in high pressure magnetron discharges

C. Arnas[1,a], A. Chami[1], L. Couëdel[1,2], T. Acsente[3], M. Cabié[4], T. Neisius[4]

[1] CNRS, Aix-Marseille université, PIIM, 13397 Marseille, France
[2] Department of Physics and Engineering Physics, University of Saskatchewan, Saskatoon SK S7N 5E2, Canada
[3] National Institute for Laser, Plasma and Radiation Physics, 077125 Magurele, Romania
[4] Aix-Marseille Université, CNRS, Centrale Marseille, FSCM, CP2M, 13397 Marseille, France

**ABSTRACT:** Nanoparticles are produced in sputtering magnetron discharges operating with a tungsten cathode at 30 Pa argon pressure. Structure analyses show that they are of core-shell type. The core is a monocrystal mainly in the metastable beta-tungsten phase and the shell is made of tungsten oxide. The origin of the metastable phase is attributed to the presence of residual oxygen in the device. Since this phase transforms into the stable alpha-tungsten phase by annealing, a standard model on the thermal balance of nanoparticles was used to find the temperature that they can reach under the considered experimental conditions. It is shown that this temperature is significantly higher than the gas one but not high enough to transform the monocrystalline metastable beta-phase during the plasma process.

--------------------------------------

[b] Author to whom correspondence should be addressed. Electronic mail: cecile.arnas@univ-amu.fr

## I. INTRODUCTION

The method of Physical Vapor Deposition (PVD) is widely used to produce deposition of thin-films of various compositions (metal, metal oxide, alloy, ceramic, polymer…) onto a broad range of substrate materials. PVD can be developed in magnetron sputtering discharges where magnetic fields are driven by permanent magnets to trap electrons near the cathode region.[1] Discharges can then be produced at low pressure due to magnetic electron confinement and the deposition process is done with energetic sputtered particles.[2] For a given magnetron discharge geometry, there is a fairly large pressure range for which a part of the slow down sputtered atoms can feed growing clusters in the plasma volume until the appearance of nanoparticles (NPs).[3] Many studies have explored their formation in plasmas for various applications extending from electronics,[4] optics,[5] magnetism,[6] catalysis,[7] to biology or medicine since the nanoscale materials present enhanced properties compared to those of bulk materials.[7,8] In particular, the NP synthesis in sputtering magnetron discharges combined with gas-aggregation systems (MS-GAS) is an efficient production way.[9-14] In discharges without magnetic field, a large number of experiments and modeling have shown that the precursors of NPs are negative molecule/cluster ions since they are trapped in the plasma potential whereas the growth of neutral an positive ions is limited due to a rapid diffusion towards the device wall.[15-17] The formation of neutral, positive and negative clusters in the presence of magnetic fields has also been evidenced with mass spectrometry[18] and might lead to the formation of NPs. It was also reported that the NP growth in MS-GAS discharges can result from nucleation in oversaturated vapor near the cathode region.[11,14]

A substantial number of experiments has been dedicated to the formation of tungsten nanoparticles (W-NPs) in sputtering discharges. An application can be found in the production of tungsten metal and alloys. It has been pointed out that grains of nanometer sizes change the parameters of production via pressing and sintering: lower temperature, shorter time to reach dense but small grain size structures.[19] The production of W-dust in the context of plasma fusion is also explored since tungsten has been selected as plasma-facing component for the next generation of tokamaks.[20] In this context, the W-dust/NP production was simulated with various plasma sources such as a quasi-stationary plasma accelerator,[21] magnetron discharges using the MS-GAS system,[13] ECR microwave discharges,[22] as well as laser and electron beam interactions with W-targets.[23,24] Basic studies on the formation of W-NPs in arc discharges,[25] sputtering discharges without magnetic field,[26,27] conventional magnetron discharges[28,29] and using the MS-GAS system have also been reported.[30] In the latter case, W-NPs were produced in two different tungsten crystal phases: the stable alpha ($\alpha$)-W phase and the metastable beta ($\beta$)-W phase.

In this paper, we report on the formation of core-shell type W-NPs. They are synthetized in conventional DC magnetron discharges operating at high argon pressure (30 Pa). The structure of the NP core is preferentially in β-W phase whereas the amorphous shell is in tungsten oxide. It is admitted that the metastable structure is stabilized with residual gas impurities such as oxygen and nitrogen[31-33] and a crystalline phase suggests that a high temperature is reached during the formation. In order to clarify the role of each parameters, we have used a standard model of the NP thermal balance. We have found that under our conditions, NPs can reach a temperature significantly higher than the gas one, this result being in agreement with previous ones obtained with NPs of other materials.[34-37] The consequences on the W-NP structure including the presence of impurities are discussed.

The description of the experimental set up and diagnostics, and the plasma parameter measurements are presented in Sec. II. Sec. III is dedicated to the NP structure analysis. The model of the NP thermal balance allowing to predict the temperature that W-NPs can reach is presented in Sec. IV. Results are discussed according to the current knowledge in this field. A conclusion is given in Sec. V.

## II. EXPERIMENTS AND DIAGNOSTICS

**II-a) Experimental set up**

A planar unbalanced DC magnetron source (mcse-ROBEKO) is used to generate metal NPs. The set-up also presented in reference 29 is schematized Fig. 1. The cathode is a W disc (99.95 purity) of 7.6 cm diameter. A grounded guard-ring of 2 cm width, 1 cm height (~ 7.4 cm inner diameter) is set under the cathode at a distance of 0.2 cm. A grounded stainless-steel disc of 15 cm diameter is placed parallel to the cathode 10 cm apart. Stainless-steel polished substrates of 1.4 cm × 1.4 cm are installed at the bottom of a compartmented rectangular box (a compartment for each substrate). The latter is handled from outside using a translating arm. During the plasma, dust particles are collected on each substrate by pushing the box under a circular hole (1.2 cm diameter) drilled at the center of the anode disc. Two half-glass cylinders, 10 cm diameter and 10 cm long are placed around the magnetron system to confine the plasma in a cylindrical geometry and to favor the NP production. In order to insert a Langmuir probe in the plasma, the half-glass cylinders are separated by two diametrically opposed gaps of ~ 1.5 cm.

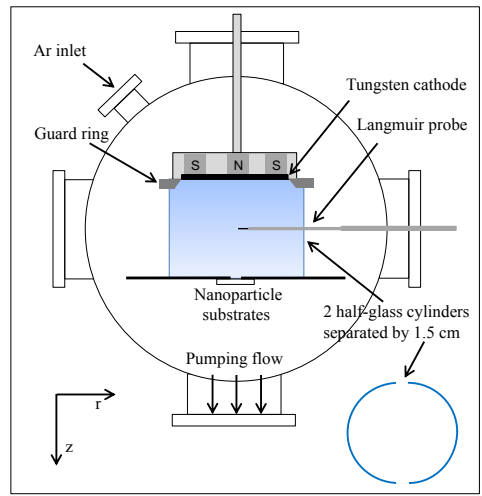

Fig. 1: Scheme of the experimental set up. The lower right insert represents the top view of the 2 half-glass cylinders. They are used to favor the trapping of NPs and opened to introduce a Langmuir probe.

The complete system is installed in a stainless steel cylinder device of 40 cm length and 30 cm diameter. The base vacuum is maintained at < $10^{-4}$ Pa using a turbo molecular pump (Edwards TurboVac 400). A DC current-regulated power supply is used (Glassman HV, 1A-1kV) to produce argon plasmas. Before each series of experiments, the cathode is cleaned by several tens of low pressure (~ $10^{-1}$ Pa), high current ( > 0.5 A) plasma pulses. After cleaning, NPs are produced in discharges generated in argon at 30 Pa under 5 sccm flow rate (standard cubic centimeters per minute) and 0.5 A current. Under such conditions, the discharge voltage applied to the cathode adjusts to ~ - 215 V (~ 100 W input power). The control of the voltage stability is made with a data acquisition card (National Instrument) at 1 kHz sampling rate.

**II-b) Langmuir probe measurements**

A Langmuir probe consisting of a 0.0125 cm diameter, 1 cm length tungsten wire is used to measure the plasma parameters radially. The probe axis is set perpendicularly to the discharge axis. To avoid its contamination, the plasma duration is limited to the duration necessary for the probe cleaning by electron bombardment (several seconds, several times) and the time required to acquire probe data, ~ 15 s. Radial measurements have been performed at a distance z = 5 cm from the cathode. At this distance, the magnetic field lines are almost parallel to the plasma axis (z direction). They are directed towards the cathode and the magnetic field measured with a Hall probe is around 3 mT radially. The Langmuir probe being displaced perpendicularly to the field lines, the effect of this weak magnetic field can be neglected during the analysis of the probe characteristics. The measured plasma parameters are then used to evaluate the temperature reached by the NPs when they are located around the plasma center.

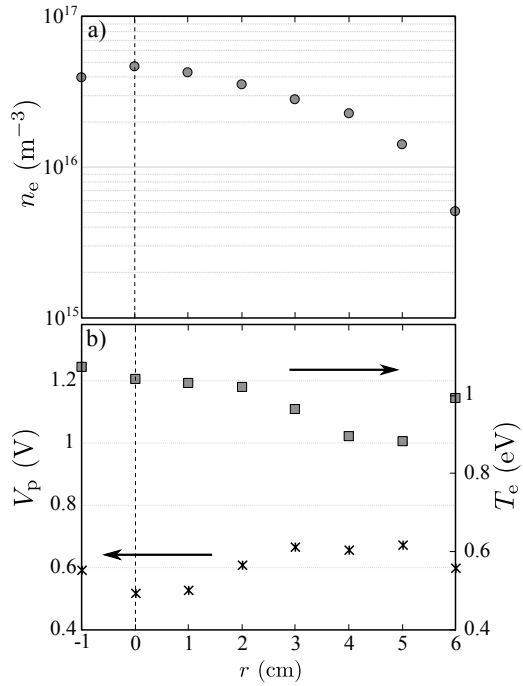

Fig. 2: At the position z = 5 cm from the cathode, radial variation of a) the electron density $n_e$ in $\log_{10}$, b) the plasma potential $V_p$ (left axis) and electron temperature $T_e$ (right axis)

Fig. 2-a) shows the radial variations of the electron density ($n_e$), and Fig. 2-b) of the electron temperature ($T_e$) and the plasma potential ($V_p$) from r = - 1 cm to r = 0 cm (plasma axis position) and from r = 0 cm to r = 6 cm (position of the half-glass cylinders). $n_e$ is maximum on the plasma axis (~ 4.5 $10^{16}$ m$^{-3}$) and decreases mainly with diffusion towards the half-glass cylinders. $T_e$ and $V_p$ varie inversely. $T_e$ increases towards the plasma edge to sustain the ionization. The average value is ~ 1 eV and the average plasma potential, ~ 0.6 V. The small average value of $T_e$ might be explained by the fact that sputtered W atoms which are not involved in the growth of NPs are ionized ($E_i$ = 7.9 eV) instead of Ar atoms ($E_i$ = 15.7 eV) in a region where electrons are no more confined and trapped as in the cathode region.

Measurements in the limit r = 6 cm may be influenced by the sheath edges of the two half glass cylinders.

**II-c) Ex-situ diagnostics for NP studies**

The morphology and size distribution of the NPs are examined by scanning electron microscopy (SEM). The images are taken by a Philips XL30 SFEG in secondary electron mode. Selected Area Electron Diffraction (SAED) is performed with a TECNAI G2 transmission electron microscope at 200 kV. A Cs corrected FEI 80-300 transmission electron microscope is used to acquire high resolution electron micrographs (HR TEM). The acceleration voltage is set at 200 kV. The instrument has a Gatan Imaging Filter (GIF Tridiem) which is used to image the oxygen distribution in the NPs by using the electrons which suffered the energy loss of ~ 532 eV related to the oxygen K-edge (Energy Filtered Transmission Electron Microscopy (EFTEM)). The background is removed by using the standard three windows method. The NPs collected on the substrates are first analyzed by SEM. Then, a portion of the NP deposit is taken with a dropper containing ethanol and put on a holey carbon TEM grid after ultrasonic dispersion.

**III. NANOPARTICLE CHARACTERISTICS AND DISCUSSION**

Fig. 3 shows a typical SEM image of NPs obtained after a magnetron discharge of 15 s, which corresponds roughly to the acquisition time of the Langmuir probe characteristic. The size distribution found from several SEM images is shown in the insert with a 3 nm bin. Fitting this distribution with a gaussian function results in an average size of ~ 21 nm and a standard deviation of 3 nm.

Details of the NPs shape, size and structure are summarized in Fig. 4. Fig. 4-a) shows NPs partly covering a hole in the amorphous carbon layer of the TEM grid. The magnification of the lower left part shown Fig. 4-b) reveals clearly the core-shell structure of the NPs. The tungsten core can have a spherical or dendritic shape. The core size varies between ~ 15 and ~ 30 nm and the thickness of the shell can be as large as 6 nm. An electron diffraction pattern of a set of NPs is shown in Fig. 4-c). The intensity of the diffractogram presented in the insert is radially averaged and analyzed. Peak positions and intensities correspond to tungsten in the β-phase.

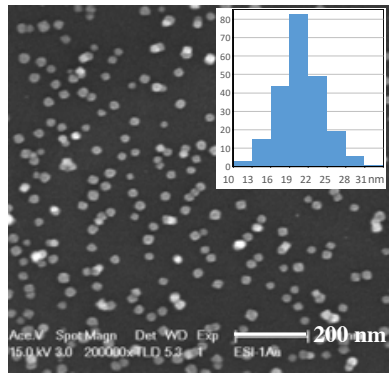

Fig. 3: SEM image of a typical W-NP sample, obtained on a stainless steel substrate. Insert: size distribution.

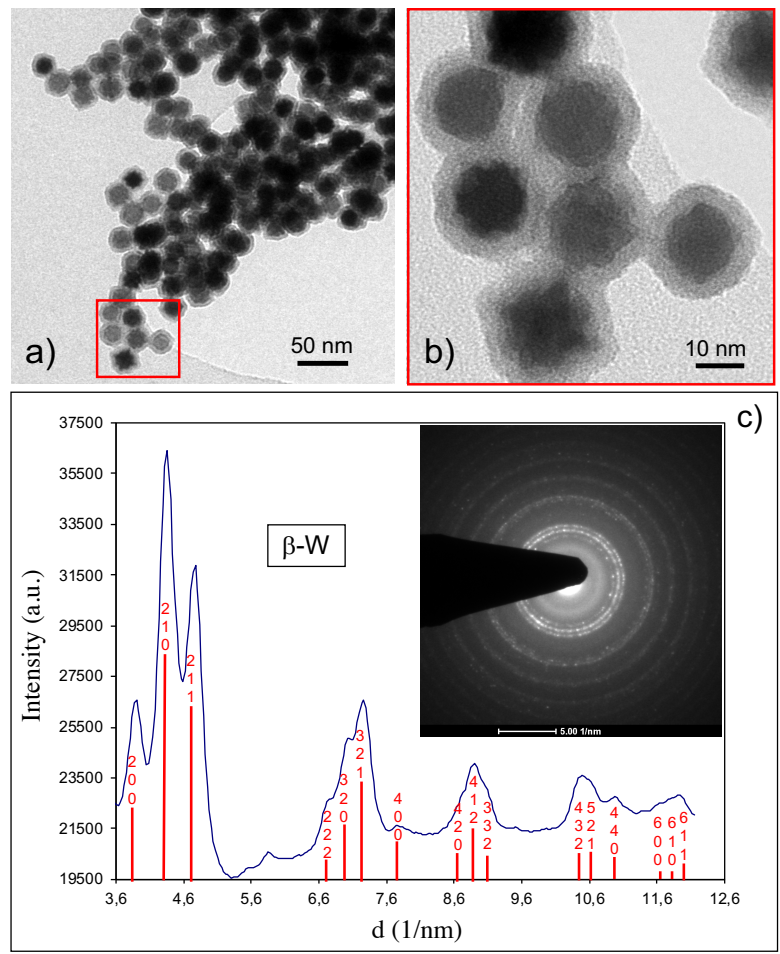

Fig.4: a) TEM image of NPs , b) magnification of the lower left part of a) in the red square, revealing core-shell type particles, c) electron diffraction pattern of a set of NPs with indices corresponding to planes of the β-phase of tungsten.

The oxygen map in Fig. 5-a) shows clearly an enrichment of oxygen in the outer shell of the NPs. This is a strong indication that the shell consists of tungsten oxide. Fig. 5-b) shows the oxygen profile of a particle on the right side of Fig. 5-a).

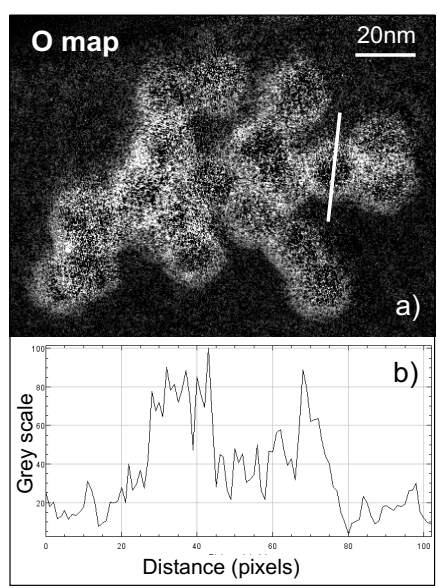

Fig. 5: a) EFTEM oxygen map showing that the NP shells contain oxygen, b) intensity profile of the oxygen content of a NP indicated by the white line in a)

From the high resolution, electron micrographs (as the ones shown in Fig. 6)) indicate that the cores are monocrystalline. Particles close to a zone axis were selected and a Fast Fourier Transform (FFT) was performed on the crystalline cores. The FFT power spectrum of these NPs close to a zone axis shown in the inserts was used to identify the core. The images in Fig. 6-a) and Fig. 6-b) show a lattice distance which corresponds to the (010) and (-110) planes of the tungsten β–phase, respectively. NPs with the α-phase were also found as presented in Fig. 6-c), where the (101) planes of the α-tungsten are identified.

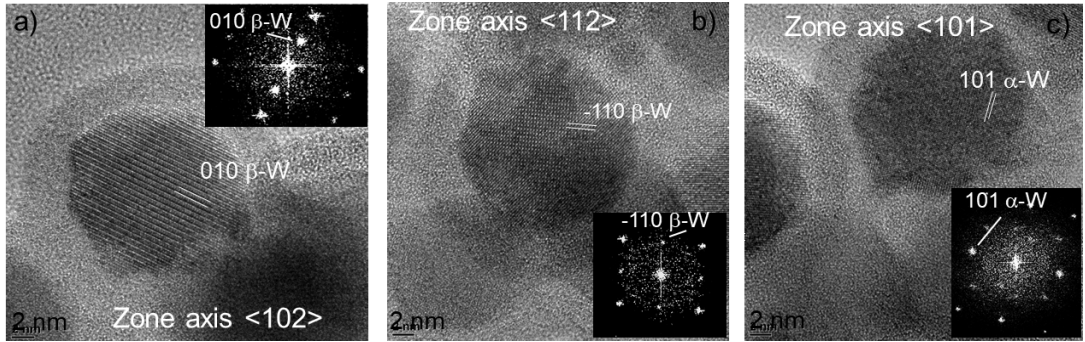

Fig. 6: HR TEM images of core-shell type NP showing that the core is monocrystalline. In a) and b) the cores are out of tungsten β-phase and in c) out of tungsten α-phase. Inserts show the FFT power spectrum of the monocrystalline cores.

The formation of W-NP in the β-phase and to a lesser extent, in the α-phase has already been observed in RF discharges equipped with the MS-GA system.[30] In such a case, the pressure was higher (80 Pa) and the input power lower (60 W). More generally, experimental studies have shown that the metastable β-phase is stabilized with gas impurities such as oxygen or nitrogen.[31-33] The presence of this phase in our conditions is consistent with the residual oxygen that always exist in discharge chambers of moderate secondary vacuum and with the use of metallic cathodes of moderate purity. In contrast, we assume that the presence of a tungsten oxide layer around the NPs is due to a long storage of the samples in the air before doing TEM and HR TEM analyses (~ 46 days). It was indeed reported in reference 13 that no tungsten oxide was visible around W-NPs when analyzes were performed one day after the device venting. Analyzes of the same sample performed one month later evidenced the presence of oxide shells.

Generally, metal NPs produced in low temperature plasmas have a crystalline structure. This also suggests a formation at high temperature. An estimation of the temperature that NPs can reach under our experimental conditions can be found using a standard model of thermal balance.

## IV. ESTIMATION OF THE NP TEMPERATURE AND DISCUSSION

The energy transfer to a NP is due to the collection of charge carriers, eventually to collisions with energetic and metastable neutrals of the background gas, the deposition of particles coming from the sputtering process and the absorption of radiation. The NP loss of energy occurs by thermal radiation and conduction with the background gas as long as its temperature remains smaller than the NP one.[34-37] Due to the thermalization of sputtered atoms of several eV average energy, the gas temperature near the cathode may indeed reach a high value.[38,39] The temperature that a NP reaches can be found at the thermal equilibrium (steady state) when the total energy flux towards the NP (influx) is equal to the heat flux from the NP (outflux).

**IV-1 Heating mechanisms**

A part of the ions Ar⁺ accelerated in the cathode sheath is reflected and neutralized at the cathode surface. At high pressure and in the plasma center region, we consider that the resulting energetic Ar atoms have already thermalized with the gaz backgroud and have no contribution to the dust heating. Collisions with metastable argon atoms (Ar$^m$) will be also neglected as well as the UV radiation absorption (no available data on Ar$^m$ density and UV radiation at high pressure). The expected effect of both mechanisms being the heating of NPs, the present study will lead to an underestimation of the NP temperature. Hence, the energy flux density $J_{in}$ towards the NP is:

$$J_{in} = j_e(\Phi)2k_BT_e + j_i(\Phi)e\Phi + j_i(\Phi)E_i + J_{chem} \qquad (1)$$

The first two terms on the right side are due to the transfer of the electron and ion kinetic energy, respectively. $j_e$ ($j_i$) is the electron (ion) flux density and $\Phi$, the NP floating potential with respect to the plasma potential. The third term is due to the ion recombination at the NP surface, $E_i$ being the ionization energy. In Eq. (1), the contribution of the tungsten ions is considered much smaller than the one of argon ions and is negligible since experiments are performed at rather low input power. For a NP of size smaller than the Debye length, $j_e$ and $j_i$ are calculated using the orbital motion-limit model. In the case of Maxwellian distributions:[40]

$$j_e = ¼\, n_0 v_e \exp(e\Phi/k_BT_e) \qquad (2)$$

$$j_i = ¼\, n_0 v_i\, (1 - e\Phi/k_BT_i) \qquad (3)$$

where $n_0$ is the plasma density and $v_e$, the electron thermal velocity. $j_i$ being expressed in the ion thermal limit, $v_i$ is the ion thermal velocity and $T_i$ the ion temperature.

The last term in Eq. (1) is due to the energy gained during the growth by tungsten deposition. Such growth is indeed suggested by the NP monocristalline structure. Assuming a stiking coeficient equal to 1, $J_{chem} = dr/dt \rho H$, where $dr/dt$ is the experimental growth rate, $\rho$ the mass density and H, the tungsten sublimation enthalpy (4,67 $10^6$ J/kg).

**IV-2 Cooling mechanisms**

The energy outflux density consists of two contributions:

$$J_{out} = J_{cond}(T_d, T_g, \alpha) + J_{rad}(a, T_d, T_g) \tag{4}$$

In low temperature plasmas, Nps of temperature $T_d$ are cooled by thermal conduction with the discharge gas ($J_{cond}$) of temperature $T_g$ and eventually, by thermal radiation emission ($J_{rad}$). The heat transfer from NPs to the discharge gas which is assumed to be only argon, is governed by Knudsen's model:[41,37]

$$J_{cond} = \frac{\gamma+1}{16(\gamma-1)} P_g \alpha \sqrt{\frac{8 k_B T_g}{\pi m_g}} \left( \frac{T_g - T_d}{T_g} \right) \tag{5}$$

where $\gamma$ is the adiabatic coefficient equal to 5/3 for argon and $m_g$, the gas atomic mass. $\alpha$ is the thermal accommodation coefficient providing information on the efficiency with which a gas exchange heat with a solid surface. For a tungsten temperature expected to be lower that 900 K, $\alpha \sim 0.3$.[42,43]

Eq. (5) which depends linearly on $T_d$, the NP temperature also depends on $T_g$. In magnetron discharges similar to ours, measurements have shown that at 30 Pa argon pressure and 100 W discharge power, the temperature under the erosion zone and under the center of the cathode are qualitatively different.[38] Under the erosion zone (the cathode center), the temperature increases (decreases) as one approaches the surface of the target. By moving away from the target, both temperatures become equal to ~ 350 K around z = 2 cm and a slow increase is measured for z > 2 cm. Extrapolation gives ~ 400 K in z ~ 4 - 5 cm from the cathode, this value being 2.5 times smaller than the one reached around z = 0.2 cm (erosion zone) where the thermalization of the sputtered atoms is at play[38]. We have also applied the model of the continuous slowing down (csd) well adapted to heavy sputtered particles and developed in reference [39]. With the csd approximation, the thermalization profiles of sputtered atoms can be established as well as the energy distribution and the percentage I(z) of particles that remain nonthermalized at a distance z from the cathode. In this model, the range of particles R(U)=Ru is used as unit path length of the energy loss per sputtered particle through collisions, U being the metal binding energy. Therefore, under our conditions of 30 Pa pressure, 100 W input power and assuming $T_g$ = 400 K, $R_u$ = 0.23 cm with U = 8.9 eV for tungsten. Taking the maximum energy of sputtered particles, $E_{max}$ = 25U (discharge voltage limit), the upper limit of the average energy of sputtered particles at the cathode surface is, ~ 3.66U = 32 eV. The average thermalization distance of the sputtered flux is then $<z_{th}>$ = $\pi R_u/2$ ~ 0.4 cm, which is very close to the cathode because of the high working pressure. The integrated probability that a particle thermalizes before reaching a plane at a distance z = 5 cm is, 1-I(5 cm) ~ 99,7%. For comparison, at z = 1 cm, 1-I(1 cm) ~ 95,0%. The latter shows that under our conditions, the sputtered atom thermalization takes place essentially at a distance smaller than 1 cm. This leads to the highest gas temperature in this area as shown with measurements in reference [38,] and from which extrapolation in z ~ 5 cm gives, $T_g$ ~ 400 K that will be used in Eq. (5).

For a particle of radius *a*, radiation emission is given by:

$$J_{rad} = \varepsilon(T_d,\lambda,a)\, P(\lambda,a)\, \Delta\lambda \qquad (6)$$

For a given radiated wavelength $\lambda$, the blackbody spectral distribution P is corrected with $\varepsilon$, the spectral emissivity of the NPs, typically smaller than the emissivity of the bulk material they are made of. The emissivity which is equal to the absorptivity for small particles, depends on the material complex refractive index. For the tungsten as for numerous metallic elements, the variations of the real and imaginary parts of the refractive index are nearly constant for radiation emission in the visible and near infrared range.[44] Since the emission of NPs takes place usually in the infrared domain in low temperature plasmas, this leads to $\varepsilon(T_d,\lambda,a) \sim \varepsilon(T_d,a)$. After integration over $\lambda$, Eq. (6) becomes:

$$J_{rad} = \varepsilon(T_d,a)\, \sigma_{SB}\, (T_d^4 - T_g^4) \qquad (7)$$

where $\sigma_{SB}$ is the constant of Stefan-Boltzman. $\varepsilon$ also varies with $T_d$, this dependency being often omitted. Precise studies were performed using the full Mie theory to obtain the absorption efficiency for a wide range of W-particle sizes and taking into account Drude theory to estimate the temperature dependent optical constants.[45] Results demonstrate strong variations of $\varepsilon$ with the temperature as a function of the radius, especially when it is smaller than 1 μm.[45,46] It was also shown that for W-NPs of radius $a \leq 10$ nm, the emissivity is smaller than $10^{-3}$ and consequently inefficient in the temperature range 300 K - 1000 K.[46] For example, when $a = 10$ nm, $\varepsilon \sim 3.5 \; 10^{-5}$ at 500 K and $\varepsilon \sim 7 \; 10^{-4}$ at 1000 K. Therefore, we have considered that the heat transfer is only due to the thermal conduction with the gas.

**IV.3 Results and discussion**

The only unknown parameter in Eq. (1) is $\Phi$. Taking the average values of the plasma parameters measured radialy in z = 5 cm and $T_g \sim 400$ K, the equality between Eq. (2) and Eq. (3) gives $\Phi \sim -3.2$ V for which, $j_e = j_i \sim 1.6 \; 10^{16}$ cm$^{-2}$s$^{-1}$. Then, the electron contribution to the NP heating is: $j_e(\Phi).2k_BT_e \sim 5.5$ mW/cm² while the total ion contribution is, $j_i(\Phi).(e\Phi + E_i) \sim 56.5$ mW/cm². Considering NPs with a W-core and a typical size of $2a \sim 15$ nm, the deposition rate is estimated around 0.5 nm/s and gives, $J_{chem} \sim 4.5$ mW/cm². Therefore, the major contribution in Eq. (1) is due by far to the ion recombination and the contribution of the incident electron flux is always smaller than the one of ions since $e\Phi > 2k_BT_e$. For a given cathode material, $J_{chem}$ depends linearly on the deposition rate. In our discharge geometry and keeping the same discharge curent, we have found that this deposition rate increases by 100% when the pressure is decreased by a factor of 2 from 40 Pa to 20 Pa.[29] In this pressure range, we may argue that the heating contribution from deposition is of the same order as the one produced with the incident electron flux.

Adding all contributions provides, $J_{in} \sim 66.5$ mW/cm². Therefore, for a W-NP of average size $2a \sim 15$ nm, $P_{in} \sim 4.7 \; 10^{-13}$ W. The equality between $P_{in}$ and $P_{out} \sim P_{cond}$ is reached when $T_d \sim 660$ K as shown Fig. 7. The radiated power of a Np of 500 K and 1000 K temperatures have been added in the figure (square and triangle, respectively). They were calculated by taking the temperature-dependent emissivity given in reference 46.

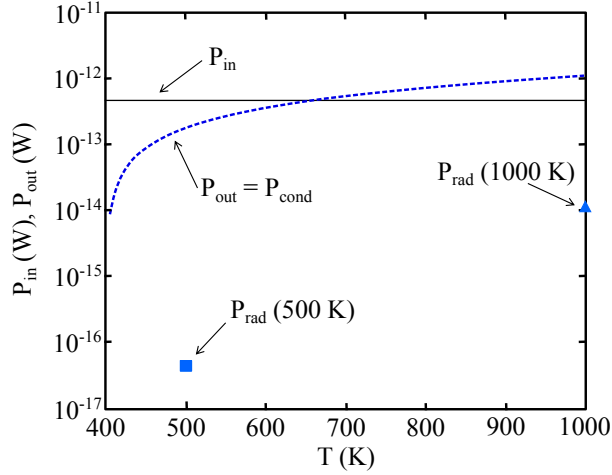

Fig. 7 : Heat power ($P_{in}$) transferred to a NP of 15 nm typical size and power transferred from this NP to the gas ($P_{out} \sim P_{cond}$). The thermal balance is achieved for a NP temperature of ~ 660 K. $P_{rad}$ (500 K)-$P_{rad}$ (1000 K) are the radiated power by a NP of temperature 500 K and 1000 K, respectively.

Their contributions are more than two orders of magnitude smaller than the one of conduction with the gas atoms.

The study of the NP thermal balance shows that even if the achieved temperature is higher than the gas discharge one, it remains lower than the W crystallization temperature which is ~ 1200 K for the tungsten bulk.[47] Similar result was already reported for aluminum, silicon NPs[34,48,49] and also carbon NPs.[36] However, Monte Carlo simulations have shown that silicon particles with *a* < 5 nm can experience highly unsteady temperature spikes, resulting from the stochasticity of exothermic events such ion recombination and surface reactions. The smaller the particle, the larger the extent of the temperature spikes.[50] In this way, the instantaneous temperature can exceed the background gas temperature by several hundreds of kelvins and reach the crystallization temperature. In the case of tungsten, the reached temperature is only evoked in the literature when studying the phase transformation from the metastable β-phase to α-phase with annealing [31,32,51] or from bombardment with reflected energetic Ar neutrals at low pressure.[33] Recent results on the W-film formation have shown that the temperature and also the transformation time depends on the oxygen impurity level, i.e., the higher the oxygen concentration, the higher the transformation temperature or the longer the duration at a given temperature for complete phase transformation.[33] In reference 52 it is reported that it is equal to 898 K for 5-12 % oxygen and consequently sufficient to activate atom diffusion, atom exchange and structure conversion. The later result suggests that the temperature reached under the plasma conditions presented in this paper cannot produce a complete conversion to the α-phase although, it remains unclear how the presence of oxygen plays into the formation of a crystalline structure.

## V. CONCLUSION

Nanoparticles of core-shell type were produced from tungsten cathode sputtering in conventional DC magnetron discharges. The operating gas was argon at high pressure relative to pressures used in PVD. Structure analyses have shown that the NP core is mainly in the metastable β-phase of tungsten while the shell is in tungsten oxide. The β-phase being stabilized with impurities, we have attributed its formation during the NP growth to the presence of residual oxygen in the device. Since this phase transforms into the stable α-phase by annealing, we have used a standard model of thermal balance to find the temperature reached by a NP under the considered experimental conditions. We have shown that the main contributions to heating is due to the ion recombination and ion flux at the surface of NPs of typical size. The cooling mechanism through conduction is not enough to lead to thermalization with the gas atoms. Hence, a temperature significantly higher than the gas one was found as shown previously for NPs of other materials but not high enough to transform the metastable β-phase during the plasma. Lastly, the presence of an oxide shell around the NPs was attributed to a long storage of samples in the air before performing structure analyses.

**Acknowledgement**: This work has been carried out within the framework of the French Federation for Magnetic Fusion Studies (FR-FCM) and the Eurofusion consortium, and has received funding from the Euratom research and training program 2014-2018 and 2019-2020 under grant agreement No 633053. The views and opinions expressed herein do not necessarily reflect those of the European Commission.